\documentclass[aps,prb,twocolumn,superscriptaddress,showpacs,amsmath,amssymb,longbibliography]{revtex4-1}
\usepackage[english]{babel}
\usepackage{amsmath}
\usepackage{bm}
\usepackage{graphicx,bbm} 
\usepackage{times}
\usepackage{epsfig} 
\usepackage[colorlinks,linkcolor=blue,citecolor=blue,urlcolor=blue]{hyperref}
\usepackage{color}
\definecolor{nred} {RGB}{224,0,0}
\definecolor{nblue} {RGB}{28,130,185}
\definecolor{dgreen} {RGB}{78,138,21}
\definecolor{norange}{RGB}{230,120,20}

\begin{document} 
%\title{Dynamical density correlations as the signature of the transition to many-body localization}
\title{Universal dynamics of density correlations at the transition to many--body localized state}
\author{M. Mierzejewski}
\affiliation{Institute of Physics, University of Silesia, 40-007 Katowice, Poland}
\author{J. Herbrych}
\affiliation{Crete Center for Quantum Complexity and Nanotechnology, Department of Physics, University of Crete, P.O. Box 2208, 71003 Heraklion, Greece}
\author{P. Prelov\v{s}ek}
\affiliation{Jo\v zef Stefan Institute, SI-1000 Ljubljana, Slovenia}
\affiliation{Faculty of Mathematics and Physics, University of
Ljubljana, SI-1000 Ljubljana, Slovenia}

%\date
\begin{abstract}
Within one--dimensional disordered models of interacting fermions we perform a numerical study of several  dynamical density correlations, which can serve as hallmarks of the transition to the many-body localized state. Results confirm that density-wave correlations exhibit quite abrupt change with the increasing disorder, with nonvanishing long-time value characteristic for nonergodic phase. In addition, our results reveal in a wide time-window a logarithmic variation of correlations in time, which we can bring in connection with the anomalous behavior of the dynamical conductivity near the transition. Our result support the view that transition to many-body localization can be characterized by universal dynamical exponents.
\end{abstract}
\maketitle
%  71.27.+a Strongly correlated electron systems; heavy fermions
%  71.30.+h	Metal-insulator transitions and other electronic transitions
%  71.10.Fd	Lattice fermion models (Hubbard model, etc.)
% 71.23.An Theories and models; localized states
%71.23.-k	Electronic structure of disordered solids
%----------------------------------------------------------------------------------------

\section{Introduction} 
%The many-body localization (MBL) has emerged in the last decade as one of the most challenging topics within the theoretical condensed-matter and statistical-physics community, being driven also by recent experiments on this phenomenon in cold gases. 
The idea of many-body localization (MBL) emerged from the well understood Anderson localization of non-interacting (NI) fermions,\cite{anderson58,mott68,kramer93,ZZZ1_1} by taking into account the many--body interaction.\cite{fleishman80,basko06} The basic claim that localization can, at large disorder, persist in the whole spectrum and consequently at all temperatures, has been by now supported by numerous studies on one-dimensional (1D) disordered models. There are essential features of the MBL phase\cite{ZZZ5_1,ZZZ5_2,ZZZ5_3,ZZZ5_4,ZZZ5_5,ZZZ5_6,ZZZ5_7,ZZZ5_8,ZZZ5_9,ZZZ5_10,ZZZ6_1} which have been confirmed numerically, predominantly for interacting spinless fermions: the change of the level statistics from a Wigner-Dyson to a Poisson-like in the nonergodic phase,\cite{oganesyan07} the vanishing of d.c. transport at any 
$T$\cite{berkelbach10,barisic10,agarwal15,gopal15,lev15,steinigeweg15,barisic16} even beyond the regime of the
linear response\cite{kozarzewski16}, the logarithmic growth of the entanglement entropy in the MBL phase,\cite{znidaric08,bardarson12,kjall14,serbyn15,luitz16,ZZZ4_1,ZZZ4_2} the nonergodic behavior of correlation functions related also with the existence of local conserved quantities.\cite{pal10,serbyn13,lev14,schreiber15,serbyn15,khemani15,luitz16,ZZZ2_1,ZZZ2_2,ZZZ2_3,ZZZ2_4,ZZZ2_5,ZZZ2_6,ZZZ2_7,ZZZ2_8,ZZZ2_9,ZZZ2_10,ZZZ2_11}

Experimental support for the MBL comes so far  from studies of cold-atom systems.\cite{schreiber15,kondov15,bordia16,boll16}  Absence of d.c. transport under constant force\cite{kondov15} and the nonergodic evolution of initially quenched state\cite{schreiber15,bordia16} have been used as the experimental criteria. In particular, in the latter studies the long-time remainder of the charge imbalance  has been used as a practical hallmark of the MBL which becomes finite within the nonergodic phase. Such a quantity is a natural counterpart of the d.c. transport quantities, as e.g., the d.c. conductivity $\sigma_0$ being nonzero in the ``normal'' ergodic phase (including also the possibility of subdiffusive transport).\cite{berkelbach10,barisic10,agarwal15,gopal15,lev15} However, the detailed theoretical and numerical analysis of such 
indicators of the MBL is still missing.\cite{schreiber15,luitz16} On the other hand, several numerical calculations of dynamical conductivity\cite{agarwal15,steinigeweg15,barisic16} as well as more general studies of dynamics using the renormalization-group approach\cite{gopal15,vosk15,potter15,luitz16} indicate that the transition is primarily characterized by dynamical critical exponent, e.g. of the dynamical conductivity $\sigma(\omega)$. 
 
In this paper, we study  dynamical density correlations within the prototype 1D model of the MBL system. We concentrate on the aspect how the Anderson localization, established for NI fermions in 1D for any disorder strength $W$, is destroyed by a modest repulsive interaction $V$. In particular, we study the time-dependent density-wave (DW) correlation functions $C(t)$, closely related to the charge 
imbalance.\cite{schreiber15,bordia16,luitz16} We show that $C(t =\infty) $  reveal quite sharp transition at large disorder $W\sim W_c$, hence they can serve as 
hallmarks of the MBL phase. 
In addition, our results show so far novel and universal logarithmic time--dependence of $C(t)$ in a very wide time--range being particularly extended near the MBL transition. Using the memory-function analysis we show that this anomalous density dynamics is closely related to the scaling of the dynamical conductivity, confirming that such universal dynamical scaling appears to be more fundamental  hallmark of the MBL than the stationary value  $C(t=\infty)$ itself. Finally, we find evidence that such anomalous dynamics is characteristic also for other disordered systems with no apparent connection with NI Anderson localization. 

\section{Model and numerical methods.} 
We consider the prototype model for the MBL, i.e. the 1D model of interacting spinless fermions with random local potentials, 
\begin{eqnarray}
H &=& - t_0 \sum_{i} \left( c^\dagger_{i+1} c_i +\mathrm{H.c.}\right) + \sum_i h_i \widetilde{n}_i+H_V\,, \label{tv} \\
H_V &=& V \sum_i \widetilde{n}_{i+1}\widetilde{n}_{i}\,, \label{hv} 
\end{eqnarray}
where $\widetilde{n}_i=n_{i}-1/2=c^\dagger_{i} c_i-1/2$. We take quenched disorder $h_i$ with a uniform distribution $-W < h_i <W$. The model~\eqref{tv} is in 1D equivalent to the anisotropic Heisenberg model with random fields.
Taking in the following $t_0=1$ as the  the energy unit, we mostly consider cases of modest interaction $V/t_0=1$ being closer to the NI case $V=0$. We also fix the density of fermions to half-filling, i.e. having $N=L/2$ fermions on a system with $L$ sites and with periodic boundary conditions. Since the MBL at larger $W$ occurs at any temperature, we study the limiting case $T \to \infty$ being the optimal one for numerical calculations.  However, in the Appendix A we show also results for finite temperatures.

The dynamical quantities are obtained mainly from exact diagonalization (ED) of the Hamiltonian, \eqref{tv}, where one can reach $L=16$. Further on we also perform calculations of the DW correlations using the microcanonical Lanczos method (MCLM),\cite{long03,prelovsek11} very suitable for the study of dynamical quantities at 
$T \gg 1$ and allowing for substantially larger system, e.g., $L=24$. In the latter method, the main restriction is finite frequency resolution of dynamical spectra, with typically $\delta \omega \sim 10^{-3}$ (for $\sim10^4$ Lanczos steps) restricting the reachable times $t < 1000$. 

\section{Density-wave correlations.} 
As a very practical tool to investigate nonergodicty in disordered system, we employ the staggered DW operator $O_s$, and its normalized autocorrelation function
\begin{eqnarray}
O_s & =&  \sum_i (-1)^i n_i, \\
C_s(t)&=& \left< \frac{\mathrm{Tr} (O_s \mathrm{e}^{iH t} O_s \mathrm{e}^{-iH t} ) } {\mathrm{Tr}( O_s^2 )} \right>. 
\label{ct}
\end{eqnarray}
$C_s(t)$ can be directly related to the time-dependent imbalance measured in the cold-atom systems.\cite{schreiber15,bordia16} Since 
$C_s(t \to \infty) > 0$ marks a nonergodic behavior, the stiffness $C_s(\infty)$ can be used as an indicator  for the MBL transition. Here,  $\langle ...\rangle$ represents average over  $ \sim 10^2 \div 10^3$
 configurations of $h_i$. 

\begin{figure}[!htb]
\includegraphics[width=0.9\columnwidth]{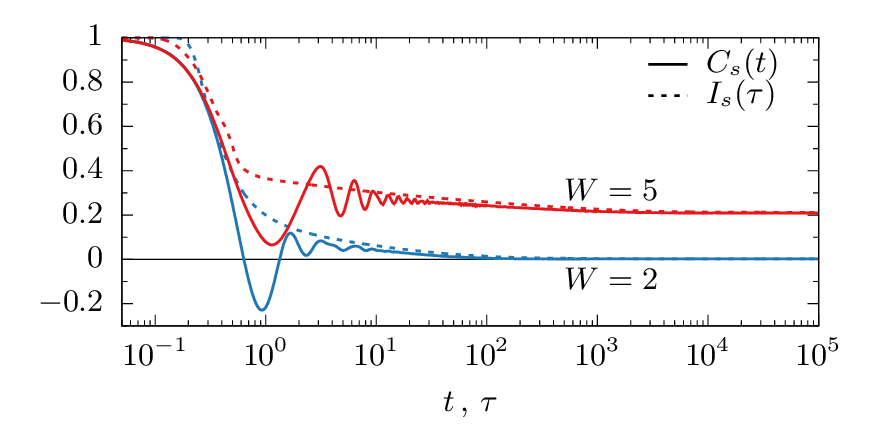}
\caption{(Color online)  Comparison of the real--time correlation function $C_s(t)$ and the integrated spectrum $I_s(\tau)$ obtained from ED for $L=16$ with weak ($W=2$) and strong ($W=5$) disorder.}
\label{fig1}
\end{figure}

The decay of $C_s(t)$ is at short $t \sim 1$ masked by oscillations with frequency $\omega=2$   which emerge from $H_0$ only.\cite{schreiber15,luitz16,kozarzewski16}
These oscillations are clearly visible in Fig. \ref{fig1}.
As discussed later on,  they can also be observed from  the Fourier transform $C_s(\omega)=1/2\pi\int^{\infty}_{-\infty} {\mathrm d}t \exp(i \omega t) \; C_s(t)$.
It is convenient to consider also a modified DW operator 
\begin{equation}
O_l = \sum_l (-1)^l \varphi^\dagger_l \varphi_l\,,
\end{equation}
as well as the corresponding correlation function $C_l(t)$. Index $l$ enumerates sorted  energies of  the single--particle Hamiltonian, 
$H-H_V= \sum_l \epsilon_l \varphi^\dagger_l \varphi_l$. 
Consequently, labelling the position of the localized states by $l$ is to some extent arbitrary. 
The correlation function $C_s(t)$  describes the decay of the initial staggered density wave with the wave--vector $q=\pi $.
Contrary  to this, $C_l(t)$ describes how/whether the system retains the information about a random  density distribution.
 Here, we search for common properties of both correlation functions which should be generic 
for most of the spatial particles distributions. Note also that the decay of $C_l(t)$ is solely due to many--body interactions,  while $C_l(t)=1$ for $V=0$.  

In order to further demonstrate universality of the long--time behavior we study also a system with a homogeneous single--particle Hamiltonian ($h_i=0$) but with a disordered many-body interaction\cite{zakrzewski}
\begin{equation}
H_V \rightarrow \tilde H_V=\sum_i 2 V_i \widetilde{n}_{i+1}\widetilde{n}_{i}+ V' \sum_i \widetilde{n}_{i+2} \widetilde{n}_{i}\,, \label{hvv}
\end{equation} 
where $V_i \ge 0$ (positive to avoid localization due to bound states) are uniformly distributed variables, $0\le V_i \le 2W $.
% In order to avoid localization due to formation of the bound states we do not consider negative $V_i$}. 
%Factor $2$ in \eqref{hvv} ensures that operators which couple to the random field have exactly the same norm as in \eqref{tv}. 
%The localization due to disordered interaction has been studied in Ref.~\onlinecite{zakrzewski} for the Bose-Hubbard model. 
%Here, we focus on the relaxation dynamics described by 
We study the autocorrelation function $C_V(t)$  for the related DW operator 
\begin{equation}
O_V =  \sum_i (-1)^i \widetilde{n}_{i+1}\widetilde{n}_{i}\,.
\end{equation}

The DW decay function $C_s(t)$ is calculated numerically using both methods described above.
$C_l(t)$   and $C_V(t)$  are obtained from ED. In the Appendix B we show for moderate time--window that $C_l(t)$    can also be obtained in a reduced basis for larger systems than accessible for ED calculations.
$C_s(t)$ and $C_l(t)$ are obtained for the Hamiltonian \eqref{tv} with $V=1$ , while $C_V(t)$ for interaction~\eqref{hvv} with $V'=1$.  Whenever our discussion applies to  all correlation functions, we omit subscripts $s,l,V$ denoting $C = C_{(s,l,V)}$ and $O = O_{(s,l,V)}$. 
%Full ED with $L \leq 16$ yields discrete spectra $C(\omega)$. 
To recognize the essential features corresponding to $t \gg 1$ (or $\omega \ll 1$) it is convenient to present the integrated spectra
\begin{eqnarray}
I(\tau) = \int\limits_{-1/\tau}^{1/\tau}\mathrm{d}\omega\, C(\omega) = \left< \frac{{\cal I}(\tau)}{{\cal I}(\tau \rightarrow 0^+)} \right>, \label{ctau}\\ 
{\cal I}(\tau)=\sum_{\alpha,\alpha'} \theta\left(\frac{1}{\tau}-|E_{\alpha}-E_{\alpha'}| \right)  |\langle \alpha |O |\alpha'\rangle|^2 , \label{how}
\end{eqnarray}
where  $H | \alpha \rangle =E_{\alpha}  |\alpha \rangle$.
When carrying out the ED calculations we first obtain  ${\cal I}(\tau)$ from Eq. (\ref{how})  and then $C(\omega)$ by differentiating Eq. (\ref{ctau}).
Neglecting accidental degeneracies,  $C(t \rightarrow \infty)$  can formally be obtained from Eqs. (\ref{ctau}) and (\ref{how}) by  restricting summation to diagonal terms $\alpha=\alpha'$.
In Figure \ref{fig1} we compare the  real--time correlation function  $C_s(t)$ with the integrated spectrum $I_s(\tau)$.  Note that $ I(\tau)$ represent to some extent the real-$t$ evolution, since $I(t)=C(t)$ in both limits $t \to 0$ and $t \to \infty$.  Since $I(\tau)$ doesn't show the transient oscillations,  it is more convenient to use the latter quantity for the studies of the long--time dynamics which turns out to be very slow.
%which are steadily decreasing functions of $\tau$ for any $L$.
% even a small one (in ED $C(\omega)$ is the sum of discrete delta functions). 

%We present in Fig.~\ref{fig2} ED results for $C(\tau)$ in systems with $L= 10, 16$, corresponding to $T \to \infty$ traced over the whole Hilbert space at half filling $N =L/2$ and sampled over $N_s \sim 100$ configurations (?). On the same Fig.~\ref{fig2}a (?) we show also the corresponding results for $L=20,24$ obtained via MCLM \cite{long03,prelovsek11}, where in the latter method the restriction is in finite frequency resolution $\delta \omega \sim 10^{-3}$ (due to restricted number of Lanczos steps $N_L \sim 10^4$ resulting also in restricted times $\tau < 1000$. Figs.~\ref{fig2}b shows results for $\widetilde C(\tau)$, corresponding to modified CDW operator $\widetilde O$, as calculated with the full ED as well with the RBA obtained on $L=20$ sites and taking $N_g=5$ generations and limiting the number of states to $N_{st} \sim 40000$.

%----------------------------------------------------------------------------------------
%----------------------------------------------------------------------------------------
%----------------------------------------------------------------------------------------
\begin{figure}[!htb]
\includegraphics[width=0.9\columnwidth]{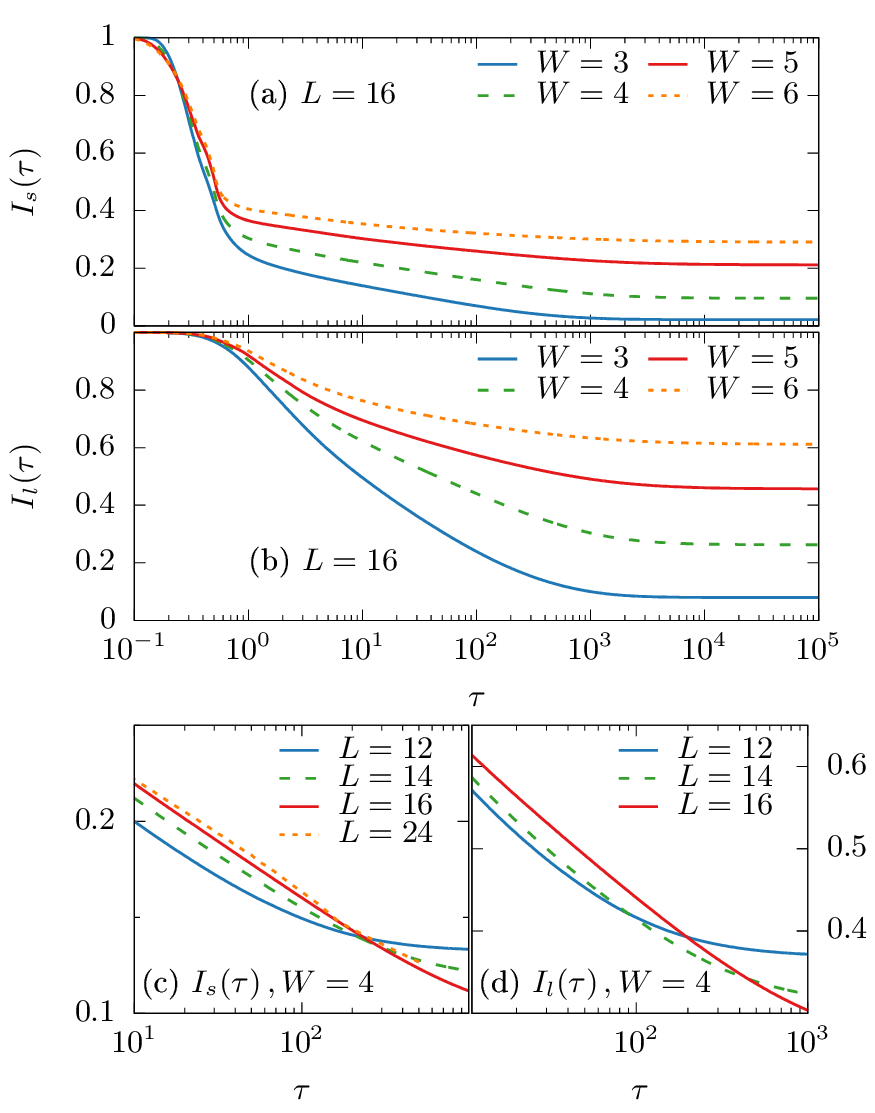}
\caption{(Color online) %a) CDW correlations $C(\tau)$, obtained via ED on systems with $L=10, 16$ sites and for different disorder $W$ at fixed interaction $V=1$. Shown are also MCLM results for $L=24$ sites, b) modified CDW correlations $C_l(\tau)$, for the same parameters, obtained with the ED as well with the RDA on $L=20$ sites. 
Integrated correlation functions $I_{s,l}(\tau)$,  obtained from ED ($L \le 16$) and MCLM ($L=24$) for different disorders $W$.}
\label{fig2}
\end{figure}
%----------------------------------------------------------------------------------------

%----------------------------------------------------------------------------------------
\begin{figure}[!htb]
\includegraphics[width=0.9\columnwidth]{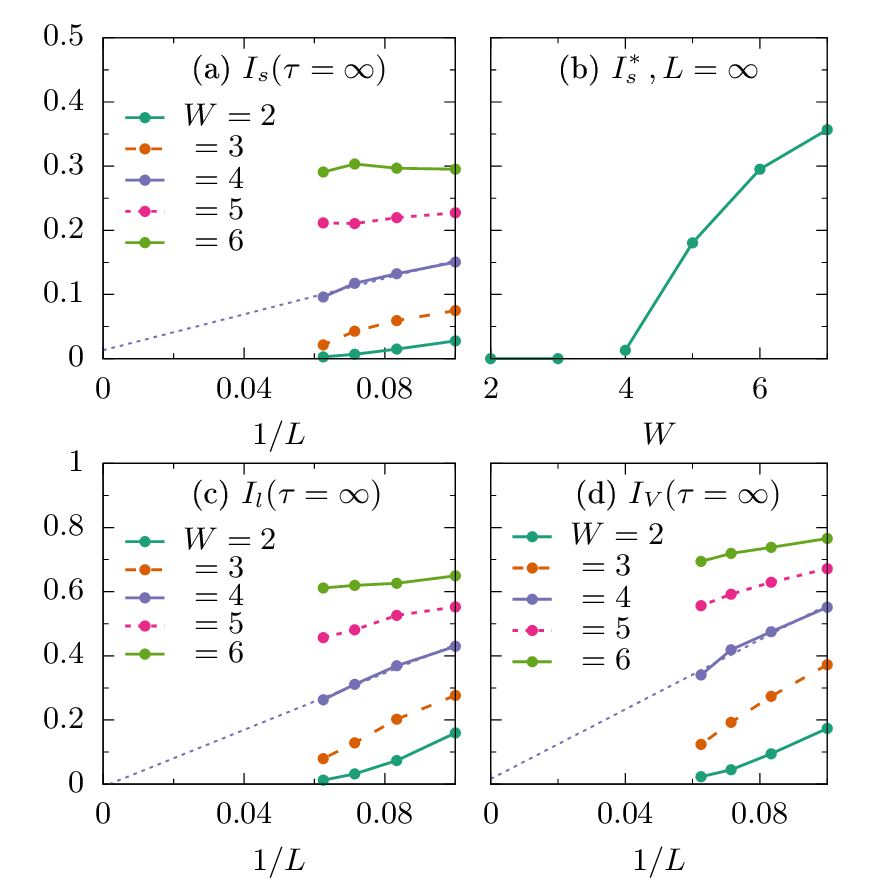}
\caption{(Color online) $1/L$ scaling of stiffnesses $I_{s,l,V}(\infty)$, obtained from ED.  Note that $C(t \rightarrow \infty)=I(\infty)$.
(b) shows the extrapolated value $I^*_s$.} 
\label{fig3}
\end{figure}
%----------------------------------------------------------------------------------------

Results for $I(\tau)$ in Figs.~\ref{fig2} and \ref{fig3}  allow for several conclusions: (a) Qualitative behavior of all  $I(\tau)$ is quite similar.
%, although $C_l(\tau)$ is smoother without a rather sharp drop visible in $C(\tau \sim 1)$ having the same origin as oscillations (or maximum at $C(\omega \sim 1)$ as reported previously\cite{schreiber15,luitz16,kozarzewski16}. 
(b) ED results at fixed $L$ always reveal finite $C(\infty)=I(\infty)>0$. These results  are plotted in Fig.~\ref{fig3} vs. $1/L$ and show that for $W<W_c \sim 4$ the scaled values vanish, in agreement with previous 
studies\cite{lev15} of the model \eqref{tv}. It is interesting that roughly the same $W_c$ is obtained for a system with the disordered interaction  \eqref{hvv}. For $W>W_c$, the extrapolation $L \rightarrow \infty$ leads to finite stiffnesses $I^*$, which can be used as convenient indicator for the MBL phase.\cite{schreiber15} (c) Most remarkable, all $I(\tau)$ reveal for $W \geq 3$ a very wide time--window beyond $\tau > 1$ with a slow, logarithmic--like decay, e.g. $I(\tau) \sim a - b \log (\tau)$.  As shown in Figs.~\ref{fig2}c and \ref{fig2}d, this behavior is particularly clear at the transition $W\simeq W_c$. 
In the latter case, deviations from the logarithmic time-dependence diminish when the system size increases, hence these deviations seem to represent the finite--size effects. 
Such a decay extends typically to $\tau^* \sim 1000$ for largest systems $L = 16$ available for ED, before saturating at $C(\infty)$. The decay continues apparently  to the largest $\tau^* \sim 1000$ available by MCLM ($L=24$). The comparison of results for different $L$ confirms that  $\tau^*$ can extend at least for one decade when increasing the system from $L=12$ to $L=16$.

\begin{figure}[!htb]
\includegraphics[width=0.9\columnwidth]{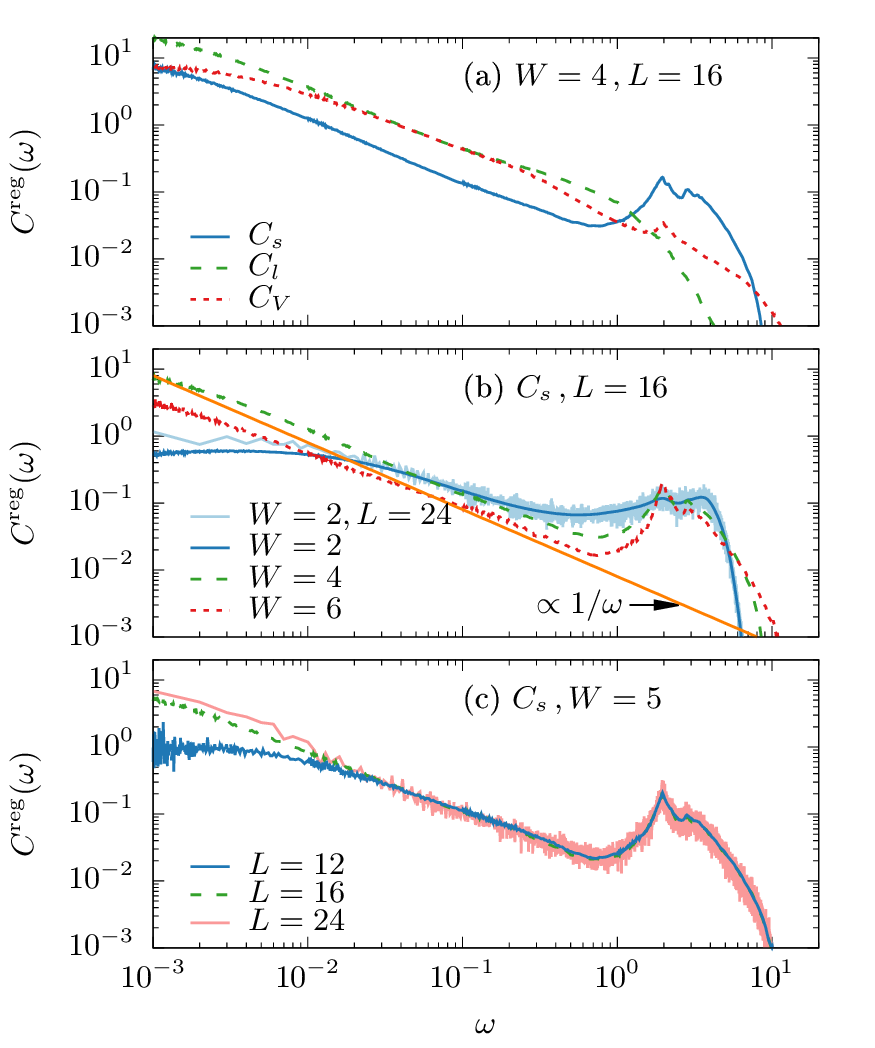}
\caption{(Color online) Regular parts of DW correlation spectra $C^{\text{reg}}(\omega)$, obtained via ED ($L \le 16$) and MCLM ($L=24$).
In (b) we include for reference the marginal behavior $C^{\text{reg}}(\omega) \propto 1/\omega$.
Note that  contrary to other correlation functions, $C_s(\omega)$ has a strong peak at $\omega=2$. } 
\label{fig4}
\end{figure}

While the observations (a) and (b) have been at least partly reported before, the universality of the slow (logarithmic) variation\cite{dyn1,dyn2} appears to be a novel one. It is evident that $I(\tau) \propto \log(\tau)$ has to emerge from an anomalous $\omega$-dependence. Taking the ED results at finite $L$ as the input, it requires 
\begin{eqnarray}
C (\omega) &=&  A \delta(\omega) + C^{\text{reg}}(\omega)\,,  \label{creg1} \\
C^{\text{reg}}(\omega \ll 1) &=& B /\bigl[ |\omega|^{\zeta}+ \Delta^{\zeta} \bigr] \,.
\label{creg}
\end{eqnarray}
Here, $A=C(t \rightarrow \infty )$ is the stiffness, while $C^{\text{reg}}(\omega)$ represents the regular part.
We note that at finite $L$ the latter is meaningful only for $\omega>\delta_0$ where $\delta_0$ is typical level spacing (with $\delta_0 \sim 10^{-2},10^{-3}$
for $L=12,16$, respectively).
We display in Fig.~\ref{fig4} the ED results for $C^{\text{reg}}(\omega)$ on a log scale, which clearly 
reveal that $\zeta \leq 1$ in a wide range of $W$ in the vicinity of $W \sim W_c$. On the other hand, the saturation with $\Delta \sim 10^{-2} > \delta_0 $ is among results well resolved only 
in the case of weak disorder, i.e. $W=2$. 
%In any finite system, the dependence~\eqref{creg} must eventually break down for sufficiently low $\omega < \Delta$, e.g., due to a finite level--spacing. 
The question remains whether $\Delta$ is finite also, e.g. for W=3. Still, our results in 
Fig.~\ref{fig4}c strongly suggest that $\Delta$  vanishes  within the MBL phase, i.e. for $W>W_c$, while more detailed behavior of $\Delta$ in the regime $W \lesssim W_c$ remains an open problem.
A strict logarithmic dependence of $C(t)$ implies that  $\zeta \rightarrow 1$, whereas Fig.  \ref{fig4} shows
that $\zeta$ is  close to but smaller than 1 leading to time dependence as $t^{-\varepsilon }$, with $\varepsilon=1-\zeta >0$.  
We cannot judge whether this tiny deviations are real or show up as numerical artifacts. 
However, since $ t^{-\varepsilon } = 1-  \varepsilon \log(t)+O[ \varepsilon^2 \log(t)^2] $ such deviations may become relevant first for long times $t \sim \exp(1/ \varepsilon)$. 
%----------------------------------------------------------------------------------------
%----------------------------------------------------------------------------------------

\section{The relation to dynamical conductivity}
\begin{figure}[!htb]
\includegraphics[width=0.9\columnwidth]{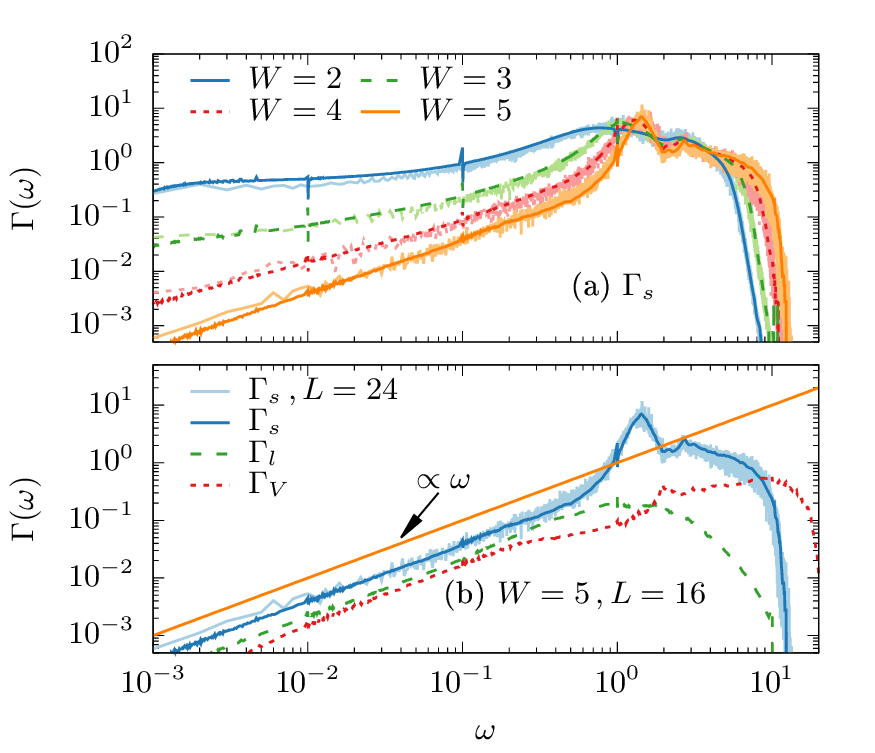}
\caption{(Color online) DW relaxation-rate functions $\Gamma(\omega)$, as calculated via the ED and MCLM  for $L=16$ and $L=24$ 
(nearly overlapping curves), respectively.  In (a)  $\Gamma_s(\omega)$ is shown for various $W$ with $L=16$ and $L=24$ , while (b) 
displays various $\Gamma(\omega)$ for  fixed $W \ge W_c$. 
%Again, for reference $\Gamma(\omega) \propto \omega$ is plotted.
}
\label{fig5}
\end{figure}

In order to relate such a behavior to other dynamical observables, it is convenient to analyze the full complex response function 
\begin{equation}
\widetilde C(\omega) = (i/\pi) \int_0^\infty\mathrm{d}t\, \exp(i\omega t)C(t),
\end{equation}
so that $\widetilde C^{\prime\prime} (\omega)=C(\omega)$. Since $\widetilde C(\omega)$ is analytical function of $\omega$ (for $ \Im \omega >0$), it can be represented in terms of the complex memory function $M(\omega)$.  
Taking into account the normalization $C(t=0)=1$ one gets
\begin{equation}
\widetilde C(\omega) = -(1/\pi) [ \omega + M(\omega)]^{-1}\,. \label{mf}
\end{equation}
In particular, for the DW correlations $C_s(\omega)$, $M(\omega)$  is related to (an effective)  
dynamical conductivity at the same $q=\pi$, i.e., $M^{\prime\prime}(\omega) \sim \sin^2(q/2) \widetilde \sigma'(q,\omega)$. 
Note, however, that  $\widetilde \sigma(q,\omega)$ is the current response function only in the limit $q\to 0$.\cite{mori65,gotze72,forster77}
In  general, the representation of $C(\omega)$, Eq.~(\ref{mf}), in terms of $M(\omega)$ has a clear advantage that instead of 
diverging $C(\omega \to 0$), we are
dealing with a regular $\Gamma(\omega) = M^{\prime\prime}(\omega)>0$, representing the DW relaxation-rate function. Results obtained for all correlation functions are compared in Fig.~\ref{fig5}. 

%----------------------------------------------------------------------------------------
%----------------------------------------------------------------------------------------

Our results indicate on two regimes with qualitatively different dynamics. For $W<W_c \sim 4$, our results are consistent with vanishing stiffness  in the thermodynamic limit, i.e., $A=0$ and only $C^{\text{reg}}$ remains in Eq.~\eqref{creg1}. Then, $\Delta$ is non-vanishing at least for weaker  disorder $W\leq 2$. Consequently, 
also $\Gamma(\omega \to 0 ) >0 $ still being much smaller then corresponding maxima appearing at $\omega  \sim 1$. 
Such $\Gamma(\omega)$ has very close similarity to  $q \to 0$ optical conductivity, $\sigma(\omega)$,\cite{barisic10,steinigeweg15,barisic16} with the form 
$\Gamma(\omega \ll 1) \sim \Gamma_0 + g \omega^{\alpha}$, where  $\alpha =\zeta \leq 1$ and $\Gamma_0 \propto \Delta^\zeta$.  
%If however $\Delta$ vanishes also in the ergodic regime, for certain $W<W_c $, then $C (\omega \ll 1)$ diverges while $\Gamma(\omega) \sim \omega^{\zeta}$.

For $W > W_c $ the nonergodic contribution $A>0$ is unavoidable. The latter leads to $\Gamma(\omega) \sim \omega^\alpha$ with $\alpha = 2- \zeta$. It is evident that the marginal case corresponds to $\zeta =1$ and $\alpha=1$, which is another criterion for the MBL transition. Our results (not shown) indicate that for $W \gg W_c$ the critical exponent is increasing,  $\alpha >1$,  nevertheless we find  in a broad range of $W$ nearly constant $\alpha \sim 1$.

\section{Conclusions} 
We presented numerical results for several dynamical density correlations. We have shown that the correlation functions
 display universal long-time behavior. It holds true for the prototype MBL Hamiltonian as well as for a system with homogenous single-particle Hamiltonian but with disordered 
 many-body interaction.\cite{zakrzewski}  All quantities reveal a nonergodic behavior, well visible in the stiffnesses, $C(t \rightarrow \infty)>0$,  which remain finite even after the extrapolation of finite-size results to $L \to \infty$. In this sense, the extrapolated values $C^*$ can be used as indicators of the nonergodic MBL phase, in direct correspondence to the imbalance stiffness  in the cold-atom experiments.\cite{schreiber15,luitz16}

Still, the main message of our study is that  all the correlation functions $C(t)$  exhibit near the MBL transition anomalously slow  relaxation towards the presumable $t \to \infty$ limit. Such a logarithmic time-dependence is visible over several decades in the window $1<t<t^*$ where $t^* > 1000$ seems to be limited at $W \geq W_c$ only by finite size restrictions of our numerical methods. Although we are dealing in our study with DW at wave--vectors $q \gg 0$, our analysis reveals a close similarity of the relaxation functions $\Gamma(\omega)$ to the behavior of the optical conductivity $\sigma(\omega)$ which is the $q \to 0$ property. The observed low-$\omega$ behavior $\Gamma \sim \Gamma_0 + g \omega^\alpha$ indicates that the MBL transition is best characterized by the critical exponent $\alpha=1$, consistent with several other numerical and renormalization-group analysis of dynamical quantities.\cite{agarwal15,gopal15,steinigeweg15,potter15,barisic16}
The extremely slow relaxation together with finite time-windows, both in
experiments as well in numerical studies, suggest that identifying MBL from the stiffness (i.e. from the saturated correlation functions)  might be very challenging.   
Hence a more proper definition of the MBL should be related just to the critical dynamics,  since this dynamics can be established within much shorter time--windows.

There are nevertheless clear open questions, e.g., whether $\Gamma_0>0$  for arbitrary  $W<W_c$  (as well a d.c. $\sigma_0 >0$)\cite{barisic10,steinigeweg15,barisic16} or  there exists intermediate regime with subdifussive behavior\cite{agarwal15,gopal15,znidaric16} consistent with $\Gamma_0=0$. 
%It has recently been shown for a classical Brownian particle that even longest--living subdiffusive regimes even may eventually be replaced by a normal diffusion  \cite{spiechowicz} .  
Beyond the fundamental importance, even more relevant is the question, whether experimental results on the imbalance relaxation 
could also confirm logarithmic character of  relaxation towards the limiting nonergodic stiffnesses.\cite{schreiber15,bordia16} 
  
\acknowledgements
%We acknowledge fruitful discussions with Fabian Heidrich-Meisner and Jakub Zakrzewski.
M.M. acknowledges support from the 2015/19/B/ST2/02856 project of the Polish National Science Center. J.H. acknowledges the European Union program FP7-REGPOT-2012-2013-1 under grant agreement n. 316165. P.P. acknowledges the support by the program P1-0044 of the Slovenian Research Agency and of the Alexander von Humboldt Foundation, which allowed for the stay at the A. Sommerfeld Center for the Theoretical Physics, LMU M\"unchen, and the Max-Planck Institute for Complex Systems, Dresden, where this work has been started. 

\appendix

\begin{figure}[!htb]
\includegraphics[width=0.9\columnwidth]{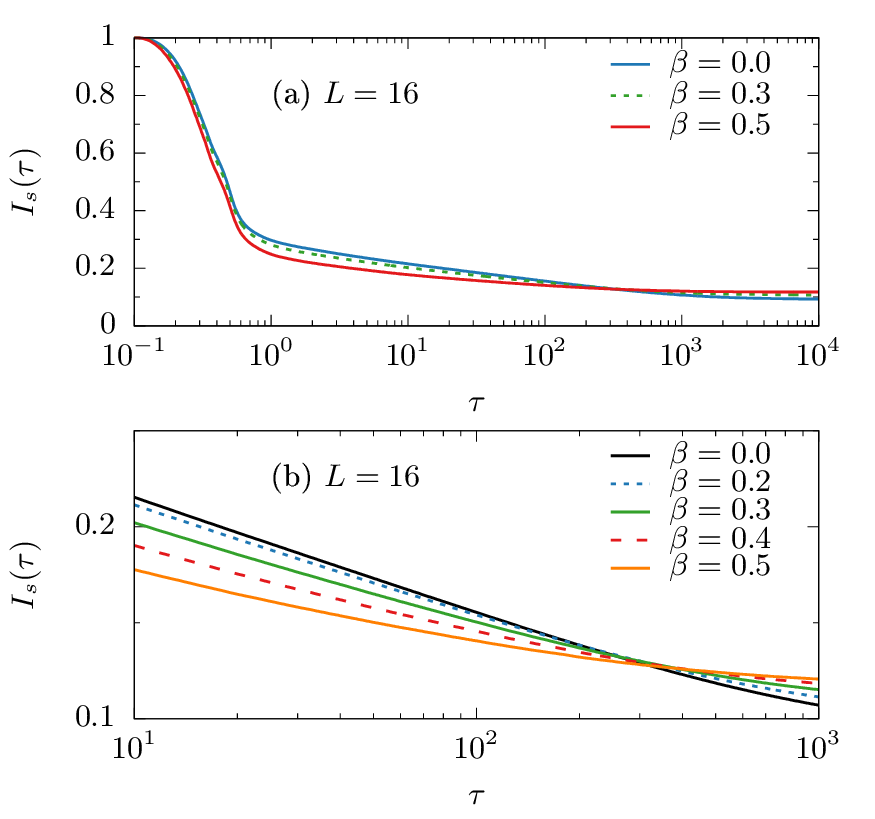}
\caption{(Color online) %a) CDW correlations $C(\tau)$, obtained via ED on systems with $L=10, 16$ sites and for different disorder $W$ at fixed interaction $V=1$. Shown are also MCLM results for $L=24$ sites, b) modified CDW correlations $C_l(\tau)$, for the same parameters, obtained with the ED as well with the RDA on $L=20$ sites. 
Integrated correlation function $I_{s}(\tau)$  obtained from exact diagonalization for $L=16$ and  $W=4$ for various inverse temperatures $\beta$.
Panel (b) shows the same as (a) but for the time--window $10< t <1000$.  }
\label{fig6}
\end{figure}

\section{Results at finite temperatures}  
All the correlation functions discussed in the main text have beed obtained at infinite temperature. While our numerical methods are inappropriate for
the studies of the low-temperature regime, these methods provide reliable results also for large but finite temperatures. Here, we demonstrate that our main conclusion concerning the dynamics of correlation functions remain valid also in the latter regime, i.e., for the state $\rho=\exp(-\beta H)/Z$, $Z=\mathrm{Tr}[\exp(-\beta H)]$ with the inverse temperature $\beta \ll 1$. 
Deep in the MBL regime one may formally assume the system to be in a thermal state.  However such assumption may be unphysical simply because  systems
do not thermalize in the latter regime.  In order to avoid this possible inconsistency we restrict our studies to the case $W=4$, i.e., we stay close to the MBL transition.  

At finite temperature, the real--time correlation function becomes
\begin{equation} 
C_s(t)= \left< \frac{\mathrm{Tr} (\rho \; O_s \mathrm{e}^{iH t} O_s \mathrm{e}^{-iH t} ) } {\mathrm{Tr}( \rho \; O^2_s )} \right>,
\end{equation} 
while the integrated spectra can be calculated from
\begin{eqnarray}
I_s(\tau) &=& \int\limits_{-1/\tau}^{1/\tau}\mathrm{d}\omega\, C_s(\omega) = \left< \frac{{\cal I}_s(\tau)}{{\cal I}_s(\tau \rightarrow 0^+)} \right>, \label{actau}\\ 
{\cal I}(\tau)&=&\sum_{\alpha,\alpha'}  e^{-\beta E_\alpha}\theta\left(\frac{1}{\tau}-|E_{\alpha}-E_{\alpha'}| \right)  |\langle \alpha |O_s |\alpha'\rangle|^2\,. \nonumber \\ \label{ahow}
\end{eqnarray}

In Fig. \ref{fig6} we show integrated correlation function,  $I_s(\tau)$,  obtained from the exact diagonalization for $L=16$. The increase of $\beta$ from $0$ to $0.5$ only weakly  affects  
$I_s(\tau)$,  as shown in figure  \ref{fig6}a.  As expected, the stifness $C(\infty)=I(\tau \rightarrow \infty)$ slightly increases with $\beta$.
Unfortunately, the limited accuracy of the finite--size scaling  of the disorder--averaged data does not allow us to judge
whether the critical disorder for the  MBL transition depends on $\beta$. However, results in Fig. \ref{fig6}  clearly show that our main claim concerning the extremely slow
quasi--logarithmic decay of correlation functions remains valid also for large but finite temperatures.

\section{The reduced basis approach}  
Since we are interested in the MBL physics emerging from the non--interacting localized Anderson states $ | l\rangle$, we introduce for comparison as well as for the closer insight the reduced basis approach (RBA). Taking $\{ | l \rangle \} $ as the basis of the single--particle space, we get 
\begin{equation}
H_0= H-H_V= \sum_l \epsilon_l \varphi^\dagger_l \varphi_l,
\end{equation}
 where $\varphi^\dagger_l = \sum_i \langle i | l \rangle c^\dagger_i $ with   $H$ and $H_V$ defined by  Eqs. (1) and (2) in the main text.  The interaction term, $H_V$, can be then written in terms of localized states as
%\begin{equation}
%H_V = V \sum_{klmn} \chi_{mn}^{kl}~\varphi^\dagger_k \varphi^\dagger_l \varphi_m \varphi_n.
%\quad \chi_{pj}^{kl} = \sum_i \phi_{ki} \phi_{l,i+1} \phi_{j,i+1} \phi_{pi}. 
%\label{hv}
%\end{equation}
\begin{equation}
H_V = \sum_{k,l,m,n} V_{klmn} \varphi^\dagger_k \varphi^\dagger_l \varphi_m \varphi_n.
\end{equation}
Considering the many--particle states within such localized basis
 \begin{equation}
 | {\underline m}\rangle = \prod_m \varphi_m^\dagger |0\rangle,
 \end{equation}
 one should separately study the diagonal part of  $H_V$  denoted  as the Hartree-Fock  term, $H_{HF}$,
 \begin{equation}
  \langle  {\underline m} | H_{HF} | {\underline n} \rangle \propto \delta_{{\underline m},{\underline n}}.
  \end{equation} 
  While $| {\underline m}\rangle$ are eigenfunction of $H_0+ H_{HF}$ with eigenvalues $E_{\underline m}$, the remaining $H'=H_V-H_{HF}$  can induce the transitions between 
different ${\underline m}$. The RBA emerges from the consideration of systems with larger disorder $W$, where $H'$ is the weakest term. 
Starting the dynamics from a chosen $| {\underline m}\rangle$, one can restrict the basis only to the states within the window $| E_{\underline n} - E_{\underline m}| < \xi V$ with $\xi \sim {\cal O}(1)$. 
%Since $H'$ connects states with up to two different single-particle states,
%One could test the convergence by taking a finite number of states generated from a particular $| {\underline m}\rangle$ by modifying only a few elements out of $\{m\}$. 
The goal is to use RBA with $N_r \ll 2^L$ basis states and evaluate within it the dynamical quantity via a direct time evolution. In this way one may bypass and even monitor the question of MB resonances, the well known problem within the theory of localization\cite{anderson58,fleishman80,vosk13,potter15,imbrie16}. We typically take $N_r \simeq 4\cdot10^4$. 

\begin{figure}[!htb]
\includegraphics[width=0.9\columnwidth]{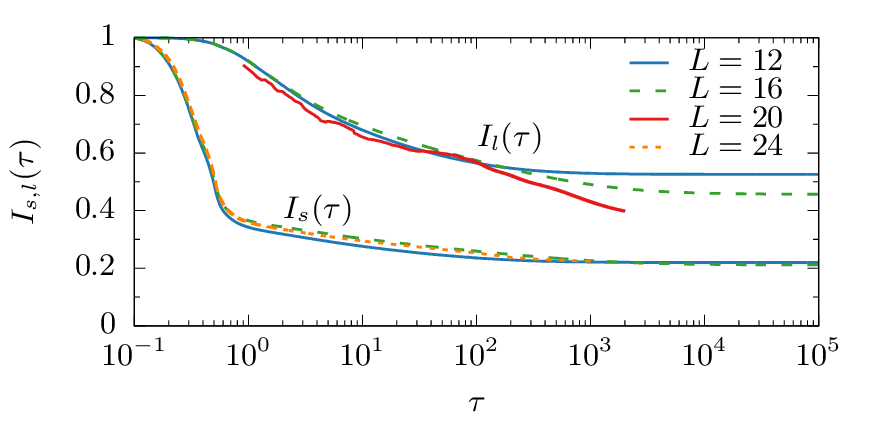}
\caption{(Color online) %a) CDW correlations $C(\tau)$, obtained via ED on systems with $L=10, 16$ sites and for different disorder $W$ at fixed interaction $V=1$. Shown are also MCLM results for $L=24$ sites, b) modified CDW correlations $C_l(\tau)$, for the same parameters, obtained with the ED as well with the RDA on $L=20$ sites. 
Integrated correlation functions $I_{s}(\tau)$ and   $I_{l}(\tau)$ obtained from  exact diagonalization ($L \le 16$), reduced basis approach ($L=20$) and microcanonical Lanczos method  ($L=24$) for disorder $W=5$.}
\label{fig7}
\end{figure}

In Fig. \ref{fig7} we compare results for the integrated correlation functions obtained from various methods, see equations (\ref{ctau}) and (\ref{how}) in the main text.  For $\tau \le 10^2$ results obtained  from exact diagonalization ($ L \le 16$) nicely overlap with the data from RBA for  $L$=20.
While, the ED results saturate for larger $\tau$,
the logarithmic decay continues apparently even further for $L=20$  up to the largest $\tau \sim 2000$ available for the RBA. The comparison of results for different $L$ confirms that  the range of logarithmic decay can extend at least for one decade when increasing
 the system from $L=10$ to $L=20$. It confirms also our main result that a more proper/practical definition of the MBL should be related to the critical dynamics rather than to stiffness. The latter quantity
 becomes available first after correlation functions saturate, while former one can be measured already during the relaxation.    

%----------------------------------------------------------------------------------------
%\bibliographystyle{apsrev4-1}
\bibliography{references}
\end{document}